\documentclass[%
 amsmath,amssymb,
 aip
]{revtex4-2}

\usepackage{bbm}
\usepackage{graphicx}
\usepackage{dcolumn}
\usepackage{bm}
\graphicspath{ {./Figures/} }

\usepackage{xcolor}

\renewcommand{\v}[1]{{{\bf #1}}}

\newcommand{\rfig}[1]{Fig.~\ref{fig:#1}}
\newcommand{\reqn}[1]{Eq.~\ref{eq:#1}}

\usepackage{amsmath}
\newcommand{\vect}[1]{\mathbf{#1}}
\renewcommand{\v}[1]{{{\bf #1}}}

\newcommand{\FT}[1]{{{\mathcal{F} \{#1\}}}}

\newcommand{\fk}[2]{e^{- i \, 2 \, \pi \, #1 \cdot #2}}
\newcommand{\fkp}[2]{e^{ i \, 2 \, \pi \, #1 \cdot #2}}

\usepackage{mathtools}

\newcommand{\vxi}{\vect{x}_i}
\newcommand{\vxo}{\vect{x}_o}
\newcommand{\vxs}{\vect{x}_s}
\newcommand{\vxsj}{\vect{x}^{(j)}_s}
\newcommand{\vxm}{\vect{x}_m}
\newcommand{\vui}{\vect{u}_i} 
\newcommand{\vuo}{\vect{u}_o}

\newcommand{\dx}{d^2 \v x}
\newcommand{\dxp}{d^2 \v x'}
\newcommand{\dui}{d^2 \vui}
\newcommand{\duo}{d^2 \vuo}
\newcommand{\dxi}{d^2 \vxi}
\newcommand{\dxo}{d^2 \vxo}
\newcommand{\dxs}{d^2 \vxs}
\newcommand{\dxm}{d^2 \vxm}

\newcommand{\Eref}{E_{\rm ref}(\vxm)}

\newcommand{\Edata}{E_3(\vxm; \vxs)}
\newcommand{\Edataj}{E_3(\vxm; \vxs^{(j)})}

\newcommand{\Esig}{E_{\rm sig}(\vxm)}
\newcommand{\Erefc}{E_{\rm ref}^*(\vxm)}
\newcommand{\Esigc}{E_{\rm sig}^*(\vxm)}
\newcommand{\Esigj}{E^{(j)}_{\rm sig}(\vxm)}
\newcommand{\Esigjo}{E^{(j)}_{\rm sig}(\vxo)}

\newcommand{\Efun}{E_1(\v x)}
\newcommand{\Efuncube}{E_1^3(\v x)}
\newcommand{\Efuncubep}{E_1^3(\vxs + \v x')}
\newcommand{\Efuncubenp}{E_1^3(\vxs + \v x)}

\newcommand{\ETHG}{E_3(\vxm)}
\newcommand{\ETHGs}{E_3(\v x - \vxs)}
\newcommand{\ETHGsj}{E_3(\v x - \vxs^{(j)})}
\newcommand{\Ein}{E_i(\vui)}

\newcommand{\Esyn}{\hat{E}_3(\vxm)}

\newcommand{\Ih}{I_{\rm h}(\vxm)}
\newcommand{\Iref}{I_{\rm ref}(\vxm)}
\newcommand{\Isig}{I_{\rm sig}(\vxm)}
\newcommand{\Iph}{\tilde{I}_{\rm h}(\vxm)}
\newcommand{\Iphc}{\tilde{I}^*_{\rm h}(\vxm)}

\newcommand{\Asig}{A_{\rm sig}(\vxm)}
\newcommand{\phisig}{\phi_{\rm sig}(\vxm)}
\newcommand{\Aref}{A_{\rm ref}(\vxm)}
\newcommand{\phiref}{\phi_{\rm ref}(\vxm)}

\newcommand{\ITHGcor}{I^c_{\rm THG}(\vxo)}

\newcommand{\Hoxm}{H_o(\vxm + \v x)}
\newcommand{\Hoxmp}{H_o(\vxo + \v x')}

\newcommand{\Po}{P_o(\vuo)}
\newcommand{\phio}{\phi_o(\vuo)}

\newcommand{\Pii}{P_i(\vui)}
\newcommand{\phii}{\phi_i(\vui)}

\newcommand{\phir}{\phi_r(\vxs)}

\newcommand{\Pop}{P_o^*(\vuo')}

\newcommand{\C}{\chi^{(3)}}
\newcommand{\CR}{\C(\v x)}
\newcommand{\CRmo}{\C(-\vxo)}
\newcommand{\CRlin}{\chi^{(1)}(\v x)}
\newcommand{\CRxs}{\C(\v x - \vxs)}

\newcommand{\CRxp}{\C(\v x')}

\newcommand{\CRs}[1]{\hat{\chi}^{(3)}(#1)}

\newcommand{\CRc}{\chi^{(3)*}(\v x')}

\newcommand{\Drr}{D(\vxm,\vxs)}
\newcommand{\Dur}{D(\vuo,\vxs)}

\newcommand{\Drrcor}{\tilde{D}(\vxm,\vxs)}

\newcommand{\Txx}{T(\vxo,\vxs)}
\newcommand{\Txxcor}{\tilde{T}(\vxo,\vxs)}

\newtagform{supplementary}[S.]()
\counterwithin*{equation}{section}

\begin{document}

\preprint{APS/123-QED}

\title{Synthetic aperture holographic third harmonic generation microscopy}

\author{Yusef Farah}
\affiliation{ 
Morgridge Institute for Research, Madison, WI 53715 USA
}%

\author{Gabe Murray}
\affiliation{%
Physics Department, Colorado State University, Fort Collins, CO 80523 USA
}%

\author{Jeff Field}
\author{Maxine Xiu}
\affiliation{%
Electrical and Computer Engineering Department, Colorado State University, Fort Collins, CO 80523 USA
}%

\author{Lang Wang}
\affiliation{ 
Morgridge Institute for Research, Madison, WI 53715 USA
}%

\author{Olivier Pinaud}%
\affiliation{%
Mathematics Department, Colorado State University, Fort Collins, CO 80523 USA
}%

\author{Randy Bartels}
\affiliation{ 
Morgridge Institute for Research, Madison, WI 53715 USA
}%
\affiliation{ 
Biomedical Engineering Department, University of Wisconsin - Madison, Madison, WI 53715 USA
}%
\email{rbartels@morgridge.org.}

\date{\today}

\begin{abstract}
Third harmonic generation (THG) provides a valuable, label-free approach to imaging biological systems. To date, THG microscopy has been performed using point scanning methods that rely on intensity measurements lacking phase information of the complex field. We report the first demonstration of THG holographic microscopy and the reconstruction of the complex THG signal field with spatial synthetic aperture imaging. Phase distortions arising from measurement-to-measurement fluctuations and imaging components cause optical aberrations in the reconstructed THG field. We have developed an aberration-correction algorithm that estimates and corrects for these phase distortions to reconstruct the spatial synthetic aperture THG field without optical aberrations.  
\end{abstract}

\maketitle


\section{\label{sec:intro}Introduction}

Nonlinear microscopy has found widespread use for imaging inside of complex scattering environments, most notably in biological tissues. Deeper imaging in these tissues is made possible by using longer wavelengths, and corresponding increases in the scattering length of light, compared with those used in linear fluorescent imaging. The nonlinear scattering process of third harmonic generation (THG) \cite{barad2003nonlinear, squier1998third} has emerged as a powerful label-free imaging modality complementary to second harmonic generation (SHG) \cite{james2021recent} and multiphoton absorption fluorescent microscopy. \cite{ carriles2009invited} THG microscopy can be used to study important cellular structures, such as lipid bodies, cell membranes, osteocytes, large protein aggregates, biogenic crystals, myelin, and microglial activation, and to differentiate subtypes of cancer. \cite{oron2004depth, debarre2006imaging, witte2011label, farrar2011vivo, lim2014label, kuzmin2016third, weigelin2016third, tokarz2017intravital, gavgiotaki2020third} The structural properties that govern the intrinsic contrast mechanism of THG are ideal for complex, heterogeneous biological samples such as a porous bone matrix and retinal layers. \cite{genthial2017label, genthial2019third, masihzadeh2015third} Furthermore, THG imaging has been shown to accurately quantify malignant tumors in breast tissue by capturing irregularities in lipid bodies that vary with microglia activation associated with different cancer subtypes and even glioma infiltration. \cite{debarre2006imaging, gavgiotaki2017distinction, zhang2019quantitative} Particularly as suitable laser systems become more readily available, THG is increasingly playing an important role in biological imaging.

Like other nonlinear microscopy methods, THG requires a sufficiently high laser light intensity to activate nonlinear signal generation due to the weak light--matter interaction strength. \cite{xu1996multiphoton, erikson2007quantification, fuentes2019second, debarre2007quantitative} To achieve sufficient signal intensity, short laser pulses are a prerequisite for nonlinear microscopy. Specifically, THG imaging is nearly always implemented in a point-scanning configuration, where a laser beam is tightly focused to a diffraction-limited point inside of a specimen. Images are built from assigning a fraction of the emitted power from each known focal point location. However, point scanning comes with several drawbacks. Traditional point-scanning multiphoton microscopes have a limited field of view (FOV). In turn, beam scanning requires an extensive amount of time to form an image with a large FOV. And as THG microscopy relies on measurements of the THG scattered power, THG microscopes are not able to capture the phase of the THG field. This results in a loss of information of the complex-valued (i.e., amplitude and phase) THG fields. To date, no THG microscopes have provided access to the amplitude and phase information of THG light scattered by a sample. The development and expansion of THG imaging methods is necessary to push the capabilities of THG microscopy for biological imaging. 

In this article, we introduce holographic imaging to directly record complex-valued THG fields, providing the first demonstration of amplitude and phase-sensitive THG holographic microscopy. This microscope is configured as a widefield, stage-scanning THG holography that enables large FOV imaging thanks to the large illumination area and stage scanning that is not limited by optical components. Use of the widefield detection scheme with a larger illumination beam than point focusing reduces the time needed for recording images, while use of a lower local illumination intensity avoids photo damaging the sample. \cite{ masihzadeh2010label, smith2013submillisecond, hu2020harmonic, fantuzzi2023wide} Further, we implement off-axis digital holography that provides heterodyne amplification to the weak THG signal and preserves the phase information of the complex-valued THG field for each illumination position.

This article also redresses a major obstacle for conventional THG imaging: that there is no recourse for correcting THG field distortions arising from aberrations that lead to poor image quality. These aberrations originate from scattered light and wavefront phase distortions introduced by imperfections in the optics and inhomogeneities in the specimen. We have developed a computational adaptive optics \cite{ adie2012computational, SMARTOCT, DISTOP, CLASS, Murray:23} approach to estimate and correct aberrations of the THG image field while preserving the signal’s phase information, allowing us to coherently combine reconstructed holograms to produce spatial synthetic aperture aberration-free THG images. We test our system on a diverse representation of samples, demonstrating its versatility. Aside from using the THG phase for aberration estimation and correction, we expect that the acquisition of phase information will prove valuable to THG three-dimensional tomography \cite{hu2020harmonic} and to study the physical origin of complex-valued nonlinear susceptibility in both biological systems and materials.


\section{\label{sec:THGholo}Third harmonic generation holography}

Our synthetic aperture THG field imaging system is built as an off-axis holographic microscope that is similar in design to SHG holography and tomography systems. \cite{masihzadeh2010label, smith2013submillisecond, hu2020harmonic, Murray:23} The experimental arrangement shown in Fig. 1 is an interferometric technique that allows us to quickly reconstruct complex images while simultaneously applying a heterodyne amplification \cite{ smith2013submillisecond} to boost weak THG signals. 

Because THG holography requires measuring interferograms with a low noise camera, it allows for a widefield imaging scenario. We record the intensity pattern formed through interference between a reference field, $\Eref$ and a signal field $\Esig$. These fields are provided by splitting the output of an ultrafast laser pulse from a Yb:fiber laser-amplifier system into a fundamental beam that illuminates the sample and a beam used to generate the reference field. The fundamental field is gently focused (to a beam diameter of $\sim 7 \mu$m, or roughly $15\times$ a typical point focus) to illuminate the sample. The signal field is generated through THG scattering produced by the quasi-widefield illumination of the sample by the fundamental beam. The THG scattered field is imaged with a 4-f microscope consisting of a UV objective lens and a 200-mm tube lens. The coherent THG reference beam is generated by tightly focusing to the back surface of a 0.5-mm thick TiO$_2$ window, where strong THG light is generated. \cite{barad2003nonlinear, squier1998third, wilson2012coherence} The THG reference light is collimated and then combined with the THG signal field thin film polarizer positioned at 45 $\deg$. The combined beam passes through a polarizer and the interference pattern is detected on an EMCCD camera. Both the signal and reference fields are at an optical frequency, $\omega_3 = 3 \, \omega_1$ that is three times the fundamental frequency $\omega_1$. These two fields interfere at the camera surface to form a holographic intensity pattern $I_h(\vxm) = |\Eref|^2 + |\Esig|^2 + \Erefc \, \Esig + \Eref \, \Esigc$. We numerically isolate the interferometric pseudo-intensity term $\tilde{I}_h = \Erefc \, \Esig$ to estimate the complex THG image field using standard protocols. \cite{ masihzadeh2010label, smith2013submillisecond} For each sample, a reference (i.e., ground truth) image is taken by illuminating the sample with a UV LED lamp to record an incoherent brightfield widefield image of the specimen to compare to the synthetic aperture THG images. A complete description of the experimental setup is detailed in the Supplemental Materials.

Each measured hologram at a sample position $\vxsj$ produces a two-dimensional (2D) complex-valued THG field map, $\Esigj$. An example set obtained by scanning $\vxs$ is illustrated in \rfig{ExperimentalConcept}. The spatial coordinates of the camera, defined by the pixel pitch, are mapped to measurement coordinates, $\vxm$, by dividing by the magnification, $M$, of the THG 4-f imaging system. This magnification was measured to be $M=47.8$ by imaging a known image test target. 

The synthetic spatial aperture is produced by scanning the specimen and recording the THG hologram for each sample position $\vxs$. Datasets of THG holograms are acquired by scanning the sample to both extend the spatial region that is imaged and to ensure sufficient overlap between adjacent hologram images, which in turn ensures convergence of our computational adaptive optics algorithm. Sufficient overlap of the hologram is obtained by translating the stage by roughly half of the illumination spot size. A stack of THG 2D field image data is formed by recording the THG field from specimen positions scanned over a large spatial aperture of the sample. This stack is then stored for data processing.

Each 2D THG image, $\Esigj$, in the data stack is flattened into a linear vector that is inserted into a column of a matrix, $\Drr$ as illustrated in \rfig{ExperimentalConcept}. The column for each specimen scan position, $\vxsj$ is set to match the lexicographical order of the $\vxm$ coordinates when the 2D field data are flattened. Because we are moving the specimen relative to the fixed THG holography image system, the matrix formed from the THG hologram data is a direct recording of the THG nonlinear distortion operator (NDO). \cite{Murray:23}

Another consequence of the fact that we move the specimen is that the illumination fundamental field, and thus the effective third-order illumination field for the THG scattering process, is the same for each measurement. Moreover, the imaging model for the collection of the THG scattered fields that is imaged onto the holography camera is also identical for each measurement. Thus, we have an intrinsically isoplanatic imaging model, provided that the imaging system for the THG scattered field is isoplanatic, which is exploited later for aberration correction.

However, a complication arises in the recording of the THG holograms. Due to random phase fluctuations arising from air density fluctuations and mechanical vibrations in the relative path between the signal and reference THG fields, there is a random phase difference between the signal and reference fields for each THG hologram measurement. Given the weak THG field strength, we require an integration time for each measured THG hologram that is long compared to the timescale for the relative phase fluctuations. If we directly integrate the hologram, the fringes will shift during the camera acquisition time, drastically reducing fringe visibility and thus degrading the THG field signal.

This phase fluctuation problem is mitigated by taking many THG holograms with short integration times for each position of the sample. Each THG field is extracted and cropped. The set of extracted fields are each flattened then stacked sequentially into columns of a matrix. Because each THG field is identical, except for a random mean phase difference between the signal and reference field and for random noise acquired on each measurement, an averaged, high signal-to-noise ratio THG field can be estimated from the dominant singular vector of the measurement matrix \cite{Murray:23}; the averaged field extracted from the dominant singular vector constitutes the measured SHG field, $\Esigj$, for the j$^{\rm th}$ sample position $\vxsj$ with a mean phase $\phi_r$.

\section{\label{sec:THGabbcorrection}THG holographic aberration correction}

\begin{figure*}
    \centering
    \includegraphics[width=0.9\textwidth]{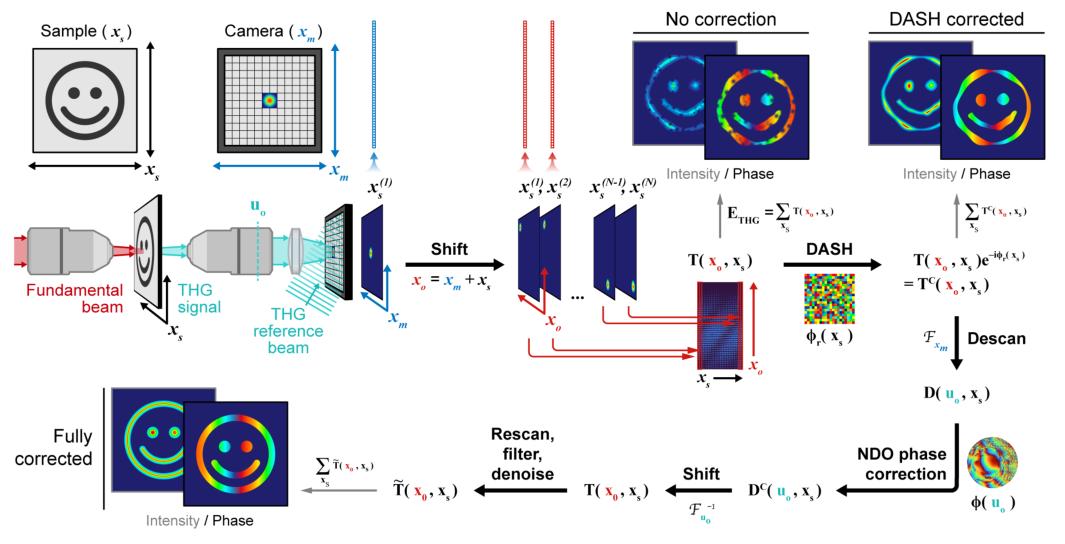}
    \caption{Diagram illustrating the collection method and aberration correction algorithm. The fundamental beam is shown in red and the generated THG is in purple. The collected THG signal and reference are overlapped on the detector to generate the hologram. Reconstructed holograms are flattened, stacked into the NDO, $\Drr$, then input to our aberration correction algorithm. In the first step of the algorithm, the THG field images are shifted, mapping $\vxm \rightarrow \vxo + \vxm$, which transforms the NDO into the transmission matrix, $\Txx$. THG field images synthesized with $\Txx$ using $\Esyn = \int \Txx \dxs$ from the raw data are highly distorted (no correction smiley face). The dominant relative phase aberration, $\phir$ is corrected with modified DASH optimization algorithm to produce a corrected transmission matrix that produces an improved THG field image (DASH corrected smiley face). Then, we further correct pupil plane aberrations with the spatial frequency-spatial NDO, $\Dur$, by estimating and correcting the pupil plane distortions $\phio$. The last step rescans the corrected spatial-spatial NDO, $\Drrcor$ to the coordinates of the corrected transmission matrix, $\Txxcor$. Spatial filtering and truncation of the singular value decomposition (SVD) of $\Txxcor$ are used to further denoise the final image. The cartoon smiley faces represent the evolution of the synthesized THG image reconstruction as the algorithm proceeds, starting with no correction, advancing to partial correction, and culminating with full correction. $\FT{\cdot}$ represents a fast Fourier transform operation that is being performed in the transformation to the $\Dur$. The summation steps that form the THG images are also followed by a reshaping step (not shown) that reverses the flattening of the image.}
    \label{fig:ExperimentalConcept}
\end{figure*}

Phase jumps between hologram measurements and pupil plane phase distortions contribute to phase aberrations that prevent the extraction of a clean THG field image. Distortions to the estimated images are illustrated by the “No correction” smiley faces in \rfig{ExperimentalConcept}. Without any corrections, the image is a poor-quality reconstruction of the THG field. Therefore, we have developed an algorithm to estimate and correct the aberrations present in the synthetic aperture THG holographic imaging process. 

The first correction in our algorithm corrects for the dominant phase distortion, $\phir$ that arises from the hologram measurement process. This phase is uniformly described over 2$\pi$, leading to speckle on an estimated image. The first step to correcting $\phir$ maps the spatial-spatial NDO to the THG transmission matrix $\Txx$. We built an optimization algorithm that maximizes the cost function, $J = \int \ITHGcor \, \dxo$, that described the power of the image, $\ITHGcor = \left| \int \Txxcor \, \dxs \right|^2$, estimated from a phase-corrected transmission matrix, $\Txxcor = \Txx \, \exp[- i \phir]$. Because aberration phase from the hologram phase difference induces a phase shift between otherwise identical complex vectors for each $\vxo$ position of the THG field, a correction of $\phir$ maximizes the estimated object intensity at each spatial point by exploiting the fact that the spatial map is highly correlated along $\vxo$. This is possible because the data shares the spatial structure of the sample nonlinear susceptibility, $\CR$. We implemented an efficient optimization algorithm that was modified from the dynamic adaptive scattering compensation holography (DASH) that was originally developed for wavefront shaping in scattering media. Details of the optimization algorithm are provided in the supplemental information. With $\phir$ correction applied to the THG field image, a clearer image can be reconstructed, depicted in the “DASH corrected” smiley faces in \rfig{ExperimentalConcept}. 

In the next step of the algorithm, we take advantage of the isoplanatic nature of the THG holographic imaging system to correct the image aberrations introduced by the pupil plane phase distortions, $\phio$, by using the concept of the distortion operator \cite{DISTOP} that is adapted for nonlinear imaging. \cite{Murray:23} To use the NDO for aberration correction, we map the “DASH corrected” spatial-spatial transmission matrix, $\Txx$, into the spatial frequency-spatial distortion operator $\Dur$. To do so, we first shift the spatial fields back from $\vxo$ to $\vxs$ to produce $\Drr$. Then each 2D output field for each $\vxs$ is Fourier transformed to the spatial frequency space, so that the NDO provides a relationship between $\vxs$ and the output spatial frequency coordinates $\vuo$. Because the THG holographic imaging system is well described by an isoplanatic model, the correction phase can be estimated from the dominate singular vector of $\Dur$. Following a similar procedure for the linear DO, \cite{ DISTOP} we take the phase and amplitude of the dominant singular vector across the output spatial frequency coordinate and apply the inverse, with a regularization parameter, to each $\vuo$ across the $\vxs$ spatial positions. Then, an inverse Fourier transform of $\vuo$ brings back the spatial-spatial NDO, $\Drrcor$, which has been aberration corrected for hologram measurement phase distortions and pupil plane phase distortions.

In the final step of our algorithm, we apply two additional noise-reduction measures to further improve the final THG field image, as depicted as the “Fully corrected” smiley faces in \rfig{ExperimentalConcept}. First, we transform the data back into the transmission matrix, $\Txx$. Taking advantage of the fact that the fundamental illumination field spans a finite spatial support, set by the width of the fundamental illumination beam, we implement a spatial filtering for each $\vxs$ location analogously to smart-OCT filtering \cite{SMARTOCT} to eliminate multiple scattering. Finally, we apply a truncated SVD and retain the singular values that contribute greater than 5\% to the total image intensity variance to further reduce noise in the image, resulting in the final corrected and complex-valued THG image, the results of which are shown in the next section. 

\section{\label{sec:level1}Results}

We applied our synthetic aperture THG imaging system and aberration correction algorithm to several samples. All reconstructed hologram images have been propagated via angular spectrum propagation to ensure images are in focus, a conventional method for hologram reconstruction. \cite{ masihzadeh2010label} Focal propagation distances are estimated using a gradient sharpness criterion.

\begin{figure*}
    \centering
    \includegraphics[width=0.9\textwidth]{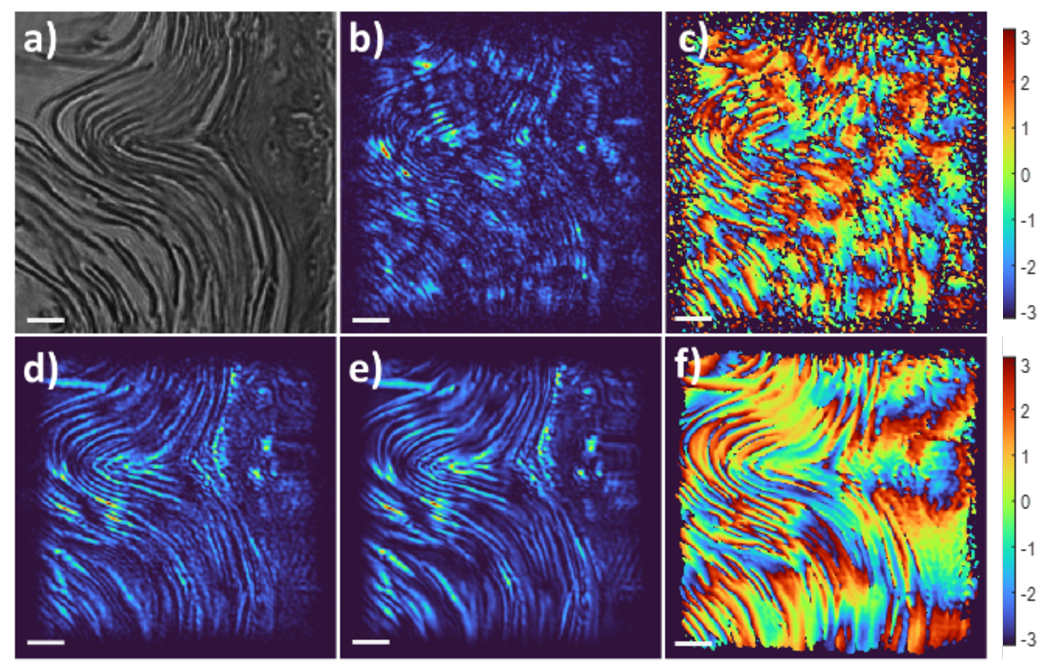}
    \caption{Mouse tail cross section. a) Brightfield image. b) THG intensity image with no correction. c) Phase map of THG image with no correction. d) Partially corrected THG intensity image with only hologram distortion phase $\phir$ corrected. e) Final THG intensity image after all corrections have been applied by our aberration correction algorithm. f) Final THG phase map with all corrections. Scale bars are 10 $\mu$m.}
    \label{fig:Mousetail}
\end{figure*}

The first set of images shown in \rfig{Mousetail} are of a $6\mu$m thick mouse tail cross section prepared on a glass slide (Triarch Incorporated). THG reconstructed images are of the epidural region of the mouse tail cross section, with the LED ground truth image shown in \rfig{Mousetail}a. The mouse tail sample demonstrates our THG holographic image reconstruction and algorithm corrections for a biological sample with a continuous structure, in this case the long and continuous strands of collagen fibers. Before any correction to the transmission matrix, the THG intensity and phase images are poor reconstructions of the mouse tail sample, as indicated in \rfig{Mousetail}b,c. After first correcting the hologram phase distortion $\phi_r(\vxs)$, the image quality improves significantly, as shown in \rfig{Mousetail}d. The fully corrected, aberration-free THG holographic images are shown in \ref{fig:Mousetail}e,f. Individual tissue strands in intensity image \rfig{Mousetail}e are well resolved, showing the fibers as continuous strands that wrap around the crimped region near the middle of the FOV. The recorded phase of individual fibers of \rfig{Mousetail}f can be identified in the phase map of the aberration-free image, reporting characteristics of sample morphology. 

\begin{figure*}
\centering
    \includegraphics[width=0.9\textwidth]{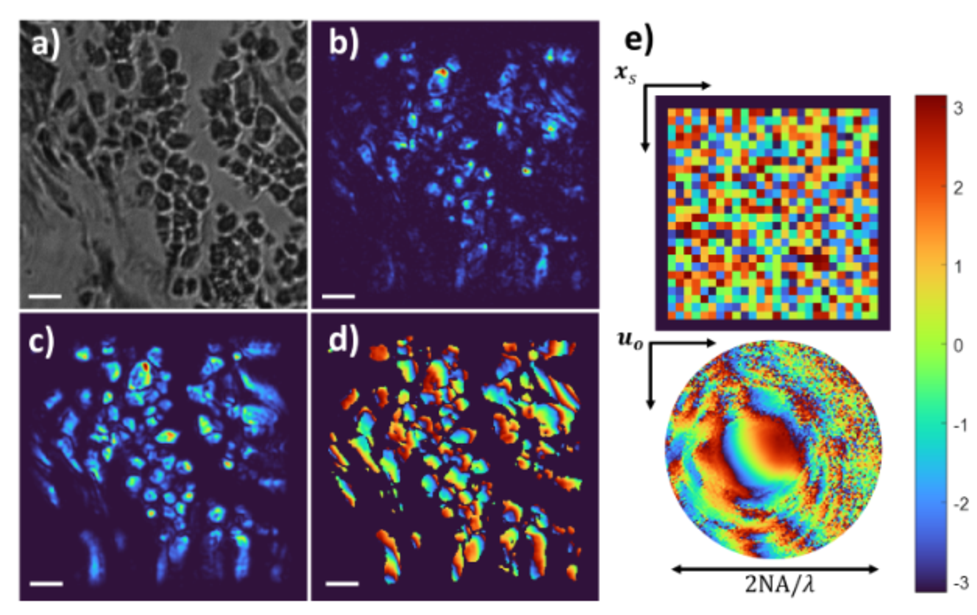}
    \caption{Developing bone. a) Brightfield image. b) THG intensity image with no correction. c) Fully corrected THG intensity image. d) Phase map of fully corrected THG image. e) (Top) The first corrections applied in our algorithm to estimate and correct $\phir$. (Bottom) NDO pupil plane correction phase map, $\phio$. Scale bars are 10 $\mu$m.}
    \label{fig:Developing_bone}
\end{figure*}

The next biological sample we demonstrate our THG holographic microscope on is a developing bone sample, approximately $6\mu$m thick (Triarch Incorporated), shown in \rfig{Developing_bone}. The THG reconstructed image shows several osteoblast cells, commonly found in developing bone. Osteoblasts are discrete cells within the bone and offer a good contrast to the continuous structure seen in the mouse tail images. When we apply our algorithm, the final THG intensity image reconstruction shown in \rfig{Developing_bone}c is of much higher quality compared to the intensity image shown in panel (b). The osteoblast cells have a discernible structure in the fully corrected THG intensity image, comparable to the brightfield image. Further, the phase map of the reconstructed field is well resolved and able to capture the signal’s phase from individual osteoblast sites. The phase corrections calculated with our algorithm are shown in \rfig{Developing_bone}e, with the top panel’s phase map showing the phase corrections applied to the transmission matrix to correct hologram phase distortions, $\phir$ and the bottom panel showing the pupil plane phase corrections acquired by the NDO step in our algorithm. 

\begin{figure*}
\centering
    \includegraphics[width=0.9\textwidth]{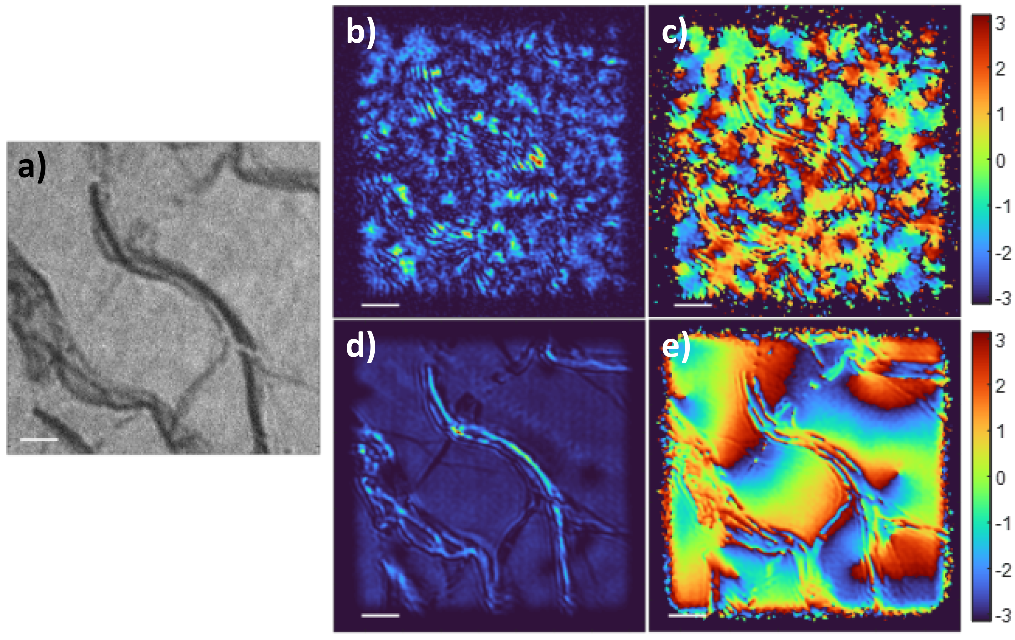}
    \caption{MoS$_2$ monolayer. a) Brightfield image. b) THG intensity image with no correction. c) Phase map of THG image with no correction. d) Fully corrected THG intensity image. e) Phase map of fully corrected THG image. Scale bars are 10 $\mu$m.}
    \label{fig:MoS2}
\end{figure*}

The final THG reconstructed image presented here is that of monolayer molybdenum disulfide (MoS$_2$) (\rfig{MoS2}). Monolayer MoS$_2$ is a two-dimensional semiconductor with a thickness of $<$1 nm and has shown significant enhancements to the third-order susceptibility through exciton resonance. In our system, the THG wavelength (355 nm) is resonant with the C-exciton and, therefore, provides a strong THG signal considering that the material is only three atoms thick. The MoS$_2$ monolayers were originally grown on a sapphire substrate via chemical vapor deposition, and then mechanically transferred to a glass microscope slide. During the mechanical transfer, wrinkles and slight defects form in the monolayer and are present in the images shown here. Similar to the biological images shown in \rfig{Mousetail} and \rfig{Developing_bone}, we see significant improvement to the reconstructed THG images after applying our aberration correction algorithm. Most interestingly, the phase map of the fully corrected image shows spatial variation of phase, even though the sample is atomically thin. 

\section{Discussion}

Here, we have demonstrated the first amplitude and phase THG images taken with the first THG holographic imaging experiment. While widefield SHG imaging and holography have been implemented by many groups, THG microscopy has remained confined to point-scanning microscopy except for a recent report on intensity widefield THG with a low-repetition rate amplifier. \cite{bernard2023third} The phase of the THG field obtained from these images provides access to a new physical measurement. We are still developing a detailed theory of THG scattering and holographic imaging to fully quantify the contributions to the THG phase, yet some conclusions can be readily drawn. 

To understand THG holographic image formation, we must understand the properties of THG scattering and how the scattered field is modified by the incident fundamental field, the spatial organization of the linear, $\CRlin$, and nonlinear, $\CR$, susceptibility, and the orientation of the harmonophores within the interaction volume. Unlike SHG, the THG scattering process is symmetry allowed in all media, meaning that THG light is always generated throughout the full focal volume. 

The differences between THG imaging with point scanning compared to widefield illumination are best illustrated by comparing the THG scattering between these two cases in a sample exhibiting a spatially uniform third-order susceptibility ($\CR$). Because ($\C$) is symmetry allowed everywhere, in a uniform medium THG is generated everywhere at a rate determined by the local strength of the electric field of the fundamental laser beam. In the case of a tightly focused beam, the scattered THG light accumulates a phase shift three times the fundamental field phase, and thus the phase of the local THG scattering depends on the fundamental field. The focused fundamental field contains the Gouy phase shift through the focus that is transferred to the THG field. Because the Gouy phase shift is an odd function about the focal plane of the beam, there is complete destructive interference of THG light scattered after the focus with light generated before the focus. Thus, in a uniform medium, no THG light is generated from a tightly focused beam. \cite{ward1969optical} Conventional laser-scanning THG imaging produces no THG signal from a homogeneous sample and only produces a signal when symmetry of the spatial distribution of the $\CR$ is broken by either an interface or a small inclusion along the direction of propagation of the beam. \cite{barad2003nonlinear} Images are formed in this configuration by recording some fraction of the scattered THG power and such measurements lose all phase information. In short, point-scanning THG imaging does not provide access to the phase of the THG field. 

By contrast, consider the case in which a plane wave fundamental beam illuminates the uniform sample. This scenario will produce a plane wave at the third harmonic frequency of the fundamental produced by THG scattering. As the phase shift accumulated with propagation by the fundamental plane wave advances linearly, there is a measurable THG field after the homogeneous sample because destructive interface across the volume is avoided. This behavior is in stark contrast to the tightly focused case. While in our experiment we are gently focusing the fundamental beam, our sample thickness is thinner than the confocal parameter of the beam, $k_1 \, w_0^2\sim 0.5$ mm, where $k_1 = n_1 \, \omega_1/c$, $n_1$ is the refractive index (RI) at the fundamental wavelength, and $w_0$ is the radius of the beam focus. Within the confocal parameter, the beam propagation can be reliably modeled as a plane wave.

Relative propagation effects between the fundamental and third harmonic field, referred to as phase matching, influence the interference of THG fields scattered at various spatial locations within the sample, thereby impacting the overall amplitude and phase of the THG field. For these two cases discussed above, we tacitly assume that there is no phase mismatch, i.e., that the RI at the fundamental and third harmonic wavelengths, $n_1$ and $n_3$, are the same. In normal materials, $\Delta n = n_3 - n_1 >0$ due to the dispersion of the RI. The collinear phase mismatch parameter is defined as $\Delta k = k_3 - 3 k_1 = \Delta n = n_3 - n_1 >0$. Here we have also written the THG wavenumber as $k_3 = n_3 \, \omega_3 /c$. The phase mismatch also affects the magnitude of the THG scattered power in both geometries, but the overall sensitivity of the THG imaging modalities to heterogeneity in $\CR$ is not substantially modified. 

Another source of phase variation in the THG holographic images occurs when the THG field propagates through the specimen’s spatial RI heterogeneity, reflected in $n_3(\v r)$, that is imprinted onto the THG field. In addition, spatial inhomogeneities in $n_1(\v r)$ can further distort the fundamental field that drives THG scattering, thereby altering the amplitude and phase of the total THG field. These two effects arising from RI inhomogeneities provide a complicated relationship between the THG field phase and the sample spatial heterogeneity in the linear and nonlinear optical susceptibility. Analysis of these effects will be the focus of future work. However, when linear scattering can be neglected, the spatial distribution of the third-order susceptibility can be estimated by using harmonic optical tomography. \cite{hu2020harmonic} 

Because the $\C$ is a tensor, the polarization of the fundamental field and the orientation of the harmonophore that generates the THG scattering also influence the complex THG image field. In many samples, the harmonophores that generate the nonlinear signals for THG are isotropically oriented, which reduces the number of THG susceptibility tensor elements and leads to a lack of angular momentum conservation for THG scattering when the specimen is illuminated with a circularly polarized beam. This conservation of angular momentum can be exploited for increased spatial resolution in THG imaging \cite{masihzadeh2009enhanced}, as can structured illumination \cite{ field2016superresolved}.  When the harmonophores are not isotopically oriented, such as with a birefringent, THG images can be recorded with circularly polarized incident light crystal. \cite{ oron2004depth, debarre2005structure, olivier2010cell}

THG phase variations can also arise from the physics of the third-order THG optical tensor. A complex-valued $\C$ will lead to phase shifts of scattered THG light that will be reflected the phase of the imaged THG field. Such a contribution to the THG phase is evident in the spatial variation of the phase in THG field image of MoS$_2$ in \rfig{MoS2}; as this sample is only three atomic layers thick, there is no possibility of phase accumulation from propagation. We speculate that the origin of this phase response arises from the spatial localization of the C-excitation in these systems, \cite{ woodward2016characterization, echarri2018enhancement} but confirmation of this conjecture requires further investigation. In general, linear and multiphoton resonances will also lead to phase variations that arise from chemical-specific absorption resonances in molecular systems. For instance, previous spectroscopy measurements of THG scattering have revealed two-photon absorption resonances in oxy- and deoxy-hemoglobin. \cite{ clay2006spectroscopy}

Because $\C$ is a weak quantity, we require an intense laser field to drive the THG scattering process sufficiently strongly to observe THG images. This weak interaction requires that we lightly focus the fundamental illumination light, even in the case of widefield imaging. The result is that our FOV is limited by practical considerations. By recording the THG field as a hologram, we benefit from heterodyne amplification of the THG field, allowing us to relax focusing conditions. However, to record images from a sufficiently large FOV that we are able to observe spatial morphologies of samples, we take a series of images as we scan the sample position. In this way, we synthesize a FOV that is not restricted by the transverse spatial extent of the fundamental illumination beam. In conventional THG imaging, stitching together a larger FOV is quite simple because we simply need to add together intensities on the boundaries. \cite{young2015pragmatic} In the case of holographic imaging, because we have access to the field information, we are extremely sensitive to phase errors when stitching together many images to synthesize a large image FOV. 

We have created an algorithm to both estimate and correct these random phase errors that inevitably occur in any experimental system to realize the synthesis of a large FOV for THG field imaging. The first step is to remove the phase jumps that occur for each measured THG field. Because these fields are recorded as a hologram, they are subject to random phase variations between measurements. This random phase noise induces phase discontinuities between different sample positions that ruin the correlations identified in the NDO. \cite{Murray:23} Without this first step, the NDO will not pick up on pupil plane phase aberrations because phase discontinuities from different sample positions hide the correlations in the NDO. The use of these phase corrections is critical for the implementation of THG holography because we need to scan the specimen to increase the FOV to observe meaningful spatial morphology; however, without phase aberration corrections, the images are too distorted to be useful. Once these aberrations are corrected, the synthetic aperture THG image field is then obtained from the corrected spatial-spatial THG transmission matrix. 

\section{Conclusion}

The results here show the incredible capabilities of our aberration-free, synthetic aperture THG holography for biological imaging. Not only have we demonstrated THG holography for the first time, but we have also developed an algorithm to reconstruct the THG field across a large FOV, preserve phase information, and correct distortions and aberrations. By scanning the specimen, we directly measure the NDO. The NDO for aberration correction is directly accessible with our experimental setup of stationary illumination. Measurements of the phase of the THG field are not accessible in conventional point-scanning THG microscopy. This phase information provides new measurement capabilities because the phase information reveals underlying resonances that introduce an imaginary component to the nonlinear optical susceptibility, i.e., $\mathrm{Im}\{ \CR \}$. Several exciting avenues are opened by this work.  By acquiring the phase of the THG signal, for example, we open a pathway for harmonic tomographic imaging with third harmonic light, while the heterodyne amplification of the THG signal opens the path for widefield THG imaging deep inside of scattering media.

\section{Acknowledgements}

We are grateful for funding support from the Chan Zuckerberg Initiative. Olivier Pinaud is supported by NSF grant DMS-2006416. Alicia Williams from the Morgridge Institute provided scientific editing support. Matt Stefely from the Morgridge Institute made Fig. 1. The MoS$_2$ sample was prepared on a glass slide and provided to us by Prof. Justin Sambur's group at Colorado State University.

\bibliography{Synthetic_aperture_holographic_third_harmonic_generation_microscopy}



\pagebreak
\widetext
\begin{center}
\textbf{\large Supplemental Materials: Synthetic aperture holographic third harmonic generation microscopy}
\end{center}
\setcounter{figure}{0}
\setcounter{table}{0}
\setcounter{page}{1}
\makeatletter
\renewcommand{\theequation}{S\arabic{equation}}
\renewcommand{\thefigure}{S\arabic{figure}}
\renewcommand{\bibnumfmt}[1]{[S#1]}
\renewcommand{\citenumfont}[1]{S#1}

\subsection{Experimental setup}

\begin{figure*}[hbt!]
    \centering
    \includegraphics[width=0.7\textwidth]{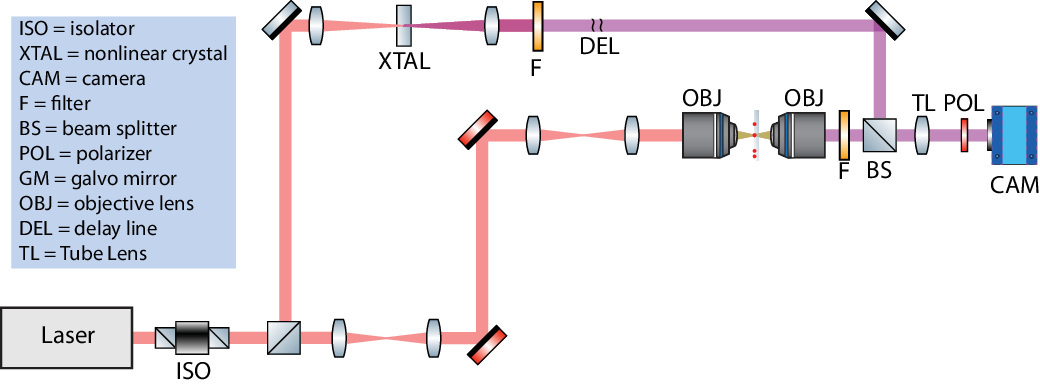}
    \caption{Experimental layout of the THG Holographic microscope for collection in a transmission geometry. The nonlinear crystal used for the reference is TiO$_2$. The red line is the fundamental light and the purple line is THG.  }
    \label{fig:Experimental_Layout}
\end{figure*}

A home-built Yb:fiber amplifier system produces $<$80 fs pulses centered at 1065 nm at a repetition rate of 40 MHz with 1 W of power. The output light from the amplifier is sent to a Martinez style compressor to pre-compensate for dispersion in the THG holographic microscope. From the compressor, the beam passes through an isolator and proceeds to the microscope section of the layout, shown in Fig. \ref{fig:Experimental_Layout}. A waveplate and polarizing beam splitter cube directs the beam into two paths, one to be used for the signal and one for the reference. The signal beam path is focused on the sample with quasi-widefield illumination by underfilling the back aperture of an aspheric objective lens (Newport, 0.5NA, 20x) with a fill ratio of 0.378. THG signal is collected in a transmission geometry by a microscope objective (Optosigma, Near UV, 0.45 NA) and reimaged to the camera (Teledyne, e2v EMCCD).

Meanwhile, after the polarizing beam splitter cube, the reference beam path is focused on the distal side of a TiO$_2$ single crystal to generate the THG reference field centered at 355 nm. The reference THG is collimated, sent through a mechanical delay stage for timing control, and is directed to the camera to interfere with the signal. A 4f configuration is used to expand the THG reference beam size to fill the camera chip and control the off-axis tilt angle relative to the signal arm. Both signal and reference arms pass through filters to select for the THG wavelength of 355 nm and polarizers to ensure similar linear polarization for optimal interference. THG holographic images are recorded using LightField® software package provided with the Teledyne camera.

\subsection{THG holography and field estimation}

The THG hologram is formed through interference between a signal field, $\Esig$, and a reference field, $\Eref$, and the intensity of that interference, where uninteresting constants of proportionality are suppressed, is given by
\begin{equation}
    \Ih = |\Eref + \Esig|^2 = \Iref + \Isig + \Iph \, e^{i \, \phi_r} + \Iphc \, e^{-i \, \phi_r}.
    \label{eq:Iholo}
\end{equation}
Here, we define the reference intensity as $\Iref=|\Eref|^2$, the signal intensity as $\Isig=|\Esig|^2$, and the pseudo intensity, from which we estimate the complex THG signal field is $\Iph =\Eref \, \Esigc$. The signal and reference fields may be separated into the amplitude, $A$, and phase, $\phi$, as $\Esig = \Asig \, \exp[i \, \phisig]$ and $\Eref = \Aref \, \exp[i \, \phiref]$. 

The intensity pattern formed during the recording of the hologram is an interferogram with phase information encoded as intensity modulations in the phase difference, $\Delta \phi = \phisig - \phiref + \phi_r$. Holography relies on the preparation of a well-known (ideally) reference field. In our case, we use off-axis holography with a uniform reference field, where the amplitude, $\Asig = A_0$ is a constant and the phase is that of a tilted plane wave, $\phiref = \v k_r \cdot \v x_m$. Here $\v k_r$ is the wavevector of the reference wave relative to the optical axis, along $\hat{z}$, the $z$-direction, of the holographic imaging system. 

We have included an additional random phase, $\phi_r$, that arises from relative phase differences accumulated in the holographic recording process. Over a set of holograms, one may see that $\phi_r$ is a random variable uniformly distributed over $[0,2 \, \pi]$. In conventional digital holography, this phase offset is not problematic, as the overall image amplitude and phase is not affected by such a uniform phase. However, in our case of synthetic spatial aperture holography, this random phase is extremely detrimental. One of the steps of our aberration correction algorithm estimates and corrects for this phase. 

\begin{figure*}[hbt!]
    \centering
    \includegraphics[width=0.6\textwidth]{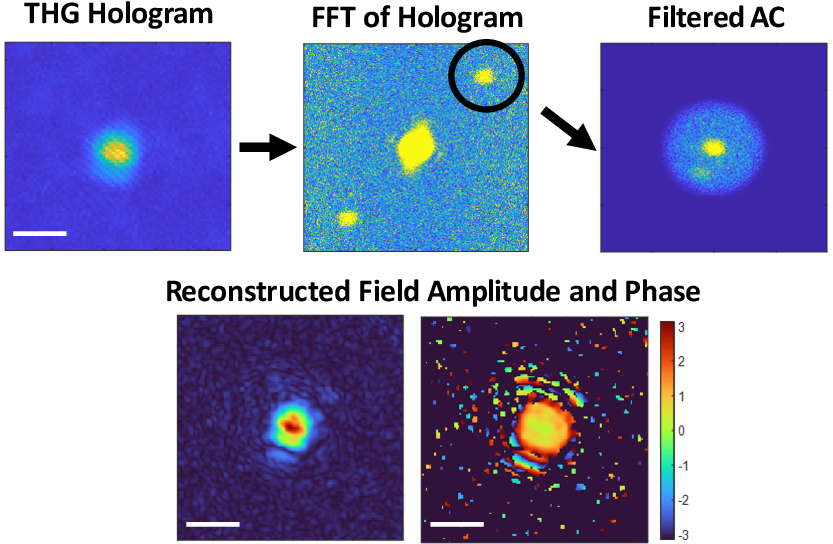}
    \caption{Reconstruction of the THG field. The hologram is a recorded interference pattern on the camera. The reconstruction is performed in MATLAB. The circular filter has a diameter of the approximate frequency support of the collection objective ($2$NA$/\lambda$). The scale bar is $7\mu$m. }
    \label{fig:Hologram_reconstruction}
\end{figure*}

With a well-characterized reference field, i.e., with a known $\Eref$, we can estimate the complex-valued signal electric field. In this case, this estimated field is that of the THG field that we have imaged to the camera. The process for estimating the THG signal field follows standard protocols \cite{masihzadeh2010label,smith2013submillisecond,hu2020harmonic} and is illustrated in \rfig{Hologram_reconstruction}. The estimation of $\Esig$ begins by taking the Fast Fourier Transform (FFT) of the THG hologram data that is described mathematically as \reqn{Iholo}, i.e., $\FT{\Ih}$. Because we have implemented off-axis holography, the terms in the FFT of \reqn{Iholo} separate into the first two terms, $\FT{\Iref}$ and $\FT{\Isig}$, appearing centered at the zero, i.e., “dc”, spatial frequency, and $\FT{\Iph}$ centered at $\v k_r$, while $\FT{\Iphc}$ is centered at $-\v k_r$.

The standard algorithm is to multiply the sideband centered at $\v k_r$ by a filter to isolate that component and compute the inverse FFT (IFFT) to recover an estimate of $\Esig$. This field is corrupted by two factors: aberrations and amplitude variations in $\Eref$ and by aberrations in the optical system that images the scattered THG field from the specimen to the camera. With our algorithm, we estimate and correct for these aberrations. As noted in the article, random fluctuations in $\phi_r$ can degrade our signal to noise ratio (SNR). SNR is boosted by rapidly acquiring a stack of THG holograms, following the THG field estimation procedure, stacking each THG field into the column of a matrix, then finally estimating the average THG field from the set of holograms from the dominate singular vector of the matrix. The phase of the dominant singular vector now contains an offset phase taken from the average of the set of THG fields for that particular scan position in the sample, $\vxsj$. This process is repeated for many sample positions, $\vxsj$'s to build a synthetic spatial aperture, enabling extended FOV synthetic aperture microscopy. 

Sample scanning is done by translating the stage in steps of 2.78 $\mu$m, equivalent to a spatial shift of 10 pixels across the camera and slightly less than half of the illumination spot size. The recorded holograms are shifted to correspond to the sample scan position, then flattened into vectors and stacked into a matrix, referred here as the transmission matrix. The columns of the transmission matrix are the different scanned positions taken across the sample plane, and the rows are the pixel positions. Adjacent widefield THG holograms within the transmission matrix have redundant spatial information because adjacent scan positions overlap. If holograms have offset phases from one another, the fields will destructively interfere with one another when the complex fields are combined.

\subsection{Imaging model and definition of the THG field transmission matrix}

The experimental setup establishes the imaging system that describes our experimental system. To properly correct for aberrations acquired in the system, we first build a model of the THG synthetic aperture imaging system that we then exploit in our aberration correction algorithm. Phase distortions from aberrations jeopardize the reconstruction of the field across the FOV. These aberrations originate from random offset phases between each hologram measurement, along with imaging system aberrations imposed by the illumination and image collection optics, and aberrations introduced by the specimen. Our goal is to estimate and correct for the sources of these phase distortions in order to produce an estimate of the complex-valued THG scattered field. To estimate these aberrations, we adapt our recently introduced method of aberration-free synthetic spatial frequency aperture SHG field imaging \cite{Murray:23} for spatially scanned extended aperture field imaging for THG.

As noted above, we obtain a scaled estimate of the scattered THG signal field, $\Esig$, that is incident on the camera. This signal field is generated through THG scattering produced by the quasi wide-field illumination of the sample by the fundamental beam, $\Efun$, that is a pulsed laser beam oscillating at a center frequency $\omega_1$. The THG field, $\ETHG$, generated in the sample region and referenced to the object plane conjugate to the camera surface, is imaged to the camera surface with a 4-f microscope imaging system, which is well described by an isoplanatic image model. 

The model for the system begins with the fundamental illumination field incident, $\Efun$, on the spatial coordinates of the object $\v x$. This fundamental field is produced by a focusing optic with an input fundamental field, $\Ein$, presented to the back focal plane of the lens with coordinates specified by the input spatial frequencies $\vui$. The illuminating fundamental field is present in the back focal plane of the illumination lens and is written as
\begin{equation}
    \Efun = \int \, \Ein \, \Pii \, \fk{\v x}{\vui} \, \dui.
\end{equation}
Here we have included the possibility of aberrations in the illumination field that are modeled as a pupil phase, $\phii$, of the illumination lens with the pupil $\Pii = \exp(i \, \phii)$, and we have assumed that the incident field underfills the input lens pupil. In addition, we have suppressed un-important scaling factors in front of the integral. We retain this suppression of unimpactful constants in all of the expressions in this derivation.

In the THG scattering process, the cube of the fundamental field drives a nonlinear source term proportional to $\CR \, \Efuncube$. For a sample with a thickness less than the phase mismatch length, $\lambda_3/(n_3-n_1)$, the image of the radiated field produced in a 4-f microscope can be written as
\begin{equation}
    \ETHG = \int  \Hoxm \, \CR \, \Efuncube \, \dx \rightarrow \Esig,
    \label{eq:THGfield}
\end{equation}
where $\vxm$ represents the spatial coordinates of the plane on the camera detector surface. The coherent spread function, $H_o(\v x)$, is the inverse Fourier transform of the complex-valued pupil function $\Po = \mathcal{X}_o(\vuo)  \, \exp(i \phio)$ in the output isoplanatic imaging pupil of the THG microscope. This pupil consists of a support defined by a circular characteristic function, $\mathcal{X}_o(\vuo)$, centered on the origin of the output spatial frequency coordinate $\vuo$ with a radius, $\mathrm{NA}/\lambda_3$, defined by the THG wavelength, $\lambda_3$, and the imaging system numerical aperture, NA. For a thin object, the THG field is directly proportional to the product $\ETHG = \CR \, \Efuncube$, and thus the desired object nonlinear susceptibility, $\CR$, may be obtained from a set of $\Esig$ measurements that are extracted from the THG hologram. this coherent spread function for the THG 4-f imaging system is given by 
\begin{equation}
    H_o(\v x) = \int \Po \, \fk{\v x}{\vuo}  \, \duo.
\end{equation}
This measured THG field represents our signal field for each measurement.

In this work, we build a synthetic aperture measurement of the specimen by taking a sequence of THG field measurements, as estimated with a holographic microscopy, for a number of measurements where the object is translated laterally by $\vxs$, leading to the set of data $\Edata = \ETHGs$. The shifted positions are a discrete set with the j$^{\rm th}$ position denoted by $\Esigj \equiv \Edataj = \ETHGsj$. Adding the shifted data to \reqn{THGfield}, we obtain the expression
\begin{equation}
    \Edata = \int  \Hoxm \, \CRxs \, \Efuncube \, \dx.
    \label{eq:THGdata}
\end{equation}
Making the variable substitution $\v x' = \v x - \vxs$, we may write
\begin{equation}
    \Edata = \int  H_o(\vxm + \vxs + \v x') \, \CRxp \, \Efuncubep \, \dxp.
\end{equation}

The fields are mapped to a synthetic spatial aperture by defining the coordinate $\vxo = \vxm + \vxsj$, so that we may write
\begin{equation}
    \Esigjo = \int  \Hoxmp \, \CRxp \, E_1^3(\vxsj + \v x') \, \dxp.
    \label{eq:THGfieldTM}
\end{equation}

Unfortunately, the extracted THG signal fields, $\Esig$, are corrupted by aberrations that manifest as phase distortions. These phase distortions arise from the measurement phase differences, $\phi_r$, and the phase distortions from the THG imaging system, $\phio$, and from the cube of the fundamental field, $\Efuncube $ that drives the THG source term in the sample. This fundamental field is also described with an isoplanatic model, $\Efun = \int \Pii \, \fk{\vui}{\vxi} \, \dxi$, with $\Pii = g(\vui) \, \exp(i \phii)$, where $g(\vui)$ is the input fundamental beam that is mildly focused to produce the quasi-widefield fundamental illumination beam and $\phii$ are pupil-plane aberrations imposed onto the illumination beam.

Estimation of these aberrations requires a set of THG fields, $\Esigj$, that exhibit spatial overlap, and thus redundancy of the underlying $\CR$ that we wish to extract. These measurements are indexed by $j$ in which the object is translated in the plane by $\vxsj$. 

\subsubsection{Transmission matrix}

The transmission matrix, $T(\vxo,\vxi)$, is an operator used in linear optics \cite{Popoff2010} that provides a map from an input field, $E_i(\vxi)$, with coordinates $\vxi$ denoting the input plane where a field is incident on an optics system. The field exiting the system, $E_o(\vxo)$, at an output plane with coordinates $\vxo$ reads
\begin{equation}
    E_o(\vxo) = \int T(\vxo,\vxi) \, E_i(\vxi) \, \dxi.
\end{equation}
In a linear system, the transmission operator for a system with a planar scattering object, described by $\chi(\v x)$, is defined by 
\begin{equation}
    T(\vxo,\vxi) = \int H_o(\vxo, \v x) \, \chi(\v x) \, H_i(\v x, \vxi) \dx.
\end{equation}
The Green's function for illumination at the input plane, $\vxi$, and propagation to the scattering plane at $\v x$, is defined as $H_i(\v x, \vxi)$. Similarly the exiting Green's function mapping the field from $\v x$ to $\vxo$ is denoted $H_o(\vxo,\v x)$. In the case of an isoplanatic (i.e., a shift invariant system), the transmission operator becomes
\begin{equation}
    T(\vxo,\vxi) = \int H_o(\vxo + \v x) \, \chi(\v x) \, H_i(\v x + \vxi) \dx.
\end{equation}

Analogously to the case of second harmonic generation holography \cite{Murray:23}, we define a transmission matrix for THG with an isoplanatic imaging system as 
\begin{equation}
    T(\vxo,\vxi) = \int H_o(\vxo + \v x) \, \CR \, H_i(\v x + \vxi) \dx.
\end{equation}
Comparing this expression to \reqn{THGfieldTM}, we see that they are identical if we make the following associations: $H_i(\v x + \vxi) \rightarrow E_1^3(\vxsj + \v x')$, $\CR \rightarrow \CRxp$, and $H_o(\vxo + \v x) \rightarrow \Hoxmp$. Now, we may write the transmission operator as 
\begin{equation}
    T(\vxo,\vxs) = \int H_o(\vxo + \v x) \, \CR \, E_1^3(\vxs + \v x) \dx,
\end{equation}
and where we note that we can exchange $\vxi$ and $\vxs$.

At this point, it is important to note that this definition of a transmission operator is not a general transmission operator as in the case of a linear transformation, but rather a sub-space of a higher-order transmission tensor. However, we can define this effective THG transmission operator that we will use for estimation of the complex THG field, which is proportional to $\CR$. Once defined over a set of the discrete spatial coordinates, the transmission operator is then represented as a transmission matrix as discussed in the article.

\subsubsection{Estimation of the synthetic THG field}

Once we have obtained a corrected form of the transmission operator matrix, we are in a position to estimate the THG field. To understand how this proceeds, we note that once the data are transformed into the transmission matrix, \reqn{THGfieldTM}, we have numerically mapped the stationary cube of the fundamental illumination field, $\Efuncube$, to an effective scanning illumination. Then, the THG source generated by this incident cubed field generates a THG field that we image with the coherent spread function $H_o(\v x)$. Let us consider an idealized case where the fundamental field is a normally-incident, unity-amplitude plane wave, i.e., $\Efun = 1$. Then, the THG transmission operator becomes
\begin{equation}
    \Txx = \int H_o(\vxo + \v x) \, \CR \dx.
\end{equation}
We see that in this approximation, we obtain a low-pass filtered image of $\CR$. In the limiting case where $H_o(\v x) \rightarrow \delta(\v x)$, then we simply have
\begin{equation}
    \Txx \approx \CRmo.
\end{equation}
However, because the illumination field has a finite spatial extent, the estimate of the THG field is obtained by integrating over the scan coordinate, $\vxs$, leading to 
\begin{equation}
    \Esyn = \int \Txx \dxs.
\end{equation}
Applying the definition of this estimator to \reqn{THGfieldTM}, we obtain
\begin{equation}
    \Esyn = \int   H_o(\vxo + \v x)  \, \CR \, F_1(\v x) \, \dx.
\end{equation}
where we have defined the synthetic illumination function
\begin{equation}
    F_1(\v x) \equiv \int E_1^3(\vxs + \v x) \, \dxs.
\end{equation}
This function gates the region of illumination of the sample. Under conditions of ideal convolution, the $F_1(\v x) \rightarrow \mathbbm{1}(\v x)$, indicating that the illumination function converges to a characteristic function that defines the region of illumination in the specimen. The goal of the aberration correction algorithm in the following section is to remove the aberrations introduced by $\Efuncube$ and $H_o(\vxo + \v x)$. With the estimated THG field, we then obtain the estimated intensity from $I_{\rm THG}(\vxo) = |\Esyn|^2$.

\subsection{Nonlinear distortion operator and aberration correction of THG spatial synthetic aperture holography}

Matrix-based approaches to optical imaging have been explored in recent years to combat multiple scattering, wavefront distortions, and aberrations in biological imaging.\cite{Popoff2010} We use the distortion operator approach inspired by Badon et. al to correct for pupil plane phase distortions.\cite{DISTOP} This approach relies on the measurement of the distortion operator, $D$. Here, we show that our experimental setup directly measures $\Drr$ in the spatial-spatial domain, and we show how $D$ is related to the transmission matrix. Here, we describe the THG nonlinear distortion operator (NDO) measured in these experiments, which is a generalization of the NDO recently introduced for SHG holographic imaging. \cite{Murray:23} In the experiments, a set of THG field data are acquired, one for each specimen location, $\vxs = \vxsj$, as given in \reqn{THGdata}. This expression is the definition of the spatial-spatial THG distortion operator, $\Drr$. As we saw in the previous section, the transmission operator arises from a coordinate shift introduced to the data, resulting in \reqn{THGfieldTM}.

In the case of an isoplanatic system, aberrations are corrected by taking a singular value decomposition of the distortion operator in the spatial-frequency/spatial space, i.e., $\Dur$. \cite{DISTOP} To obtain this mixed-space operator, we transform the output spatial coordinate to the conjugate spatial frequency coordinate, $\vuo$, using the transform
\begin{equation}
    \Dur = \int \Drr \, \fkp{\vxm}{\vuo} \, \dxm.
\end{equation}
Using the definition of the spatial-spatial NDO in \reqn{THGdata}, we obtain
\begin{equation}
    \Dur = \iint  \Hoxm \,  \fkp{\vxm}{\vuo}  \, \CRxs \, \Efuncube \, \dx \, \dxm.
\end{equation}
Making the variable substitution $\v x' = \v x - \vxs$ and using $\vxo = \vxs + \vxm$ leads to 
\begin{equation}
    \Dur = \fk{\vxs}{\vuo} \,  \int  \fkp{\vxo}{\vuo} \, \left( \int  H_o(\vxo + \v x')   \, \CRxp \, \Efuncubep \, \dxp \right) \,  \dxo.
\end{equation}
Identifying the quantity in the parentheses as the transmission operator, $\Txx$, we obtain the standard definition of the distortion operator \cite{DISTOP, Murray:23}
\begin{equation}
    \Dur = \fk{\vxs}{\vuo} \,  T(\vuo,\vxs),
\end{equation}
where we have defined the spatial-frequency/spatial transmission operator as 
\begin{equation}
    T(\vuo,\vxs) = \int  \fkp{\vxo}{\vuo} \, \Txx \,  \dxo.
\end{equation}
Applying our assumption of an isoplanatic system, we may write
\begin{equation}
    T(\vuo,\vxs) =  \Po   \, \int  \fk{\v x'}{\vuo} \CRxp \, \Efuncubep \, \dxp.
\end{equation}
We may then write the distortion operator in the isoplanatic case as
\begin{equation}
    \Dur =\Po   \, \int  \CR \, \Efuncubenp \, \fk{(\vxs + \v x)}{\vuo} \, \dx.
\end{equation}

In order to understand how and when the aberrations can be estimated, we consider the correlation
\begin{equation}
    D D^*(\vuo,\vuo') = \Po \, \Gamma(\vuo,\vuo') \, \Pop,
\end{equation}
where
\begin{equation}
    \Gamma(\vuo,\vuo')  = \iint \CR \, \CRc \, K(\vuo' - \vuo, \v x' - \v x)\,  \fk{\v x}{\vuo} \, \fkp{\v x'}{\vuo'} \, \dx \, \dxp,
\end{equation}
and with the illumination correlation function defined as
\begin{equation}
    K(\Delta \v u ,\Delta \v x) = \int E_1^3(\vxs) \, E_1^{3*}(\vxs + \Delta \v x) \, \fk{\vxs}{\Delta \v u } \, \dxs,
\end{equation}
where we have identified $\Delta \v x = \v x' - \v x$ and $\Delta \v u = \vuo-\vuo' $. The left singular vectors of the SVD of the distortion operator $D$ will then yield the eigenvectors of the correlation $D D^*$.

The form of $\Gamma$ depends on two important characteristic lengths: the spatial support of the illumination, $\ell_1$, and a typical length for $\CR$, $\ell_0$, which models either its spatial support or its scale of oscillation, whichever is shortest. There are various scenarii depending on which length is shortest, and we suppose as a start that $\ell_0$ corresponds to the spatial support of $\chi^{(3)}$. Here, $E_1^3(\vxs)$ is a function of $\vxs/\ell_1$ and  $\chi^{(3)}(\v x)$ is a function of $\v x /\ell_0$. When $\ell_0 \ll \ell_1$, then the $\Delta x$ can be ignored in $K$ and replaced by $0$, so that 
\begin{equation}
    \Gamma(\vuo,\vuo')  = \CRs{\vuo} \, K(\vuo' - \vuo,0) \,  \CRs{\vuo'}^*
\end{equation}
and then
\begin{equation}
    D D^*(\vuo,\vuo') = \Po \,  \CRs{\vuo} \, K(\vuo' - \vuo,0) \,  \CRs{\vuo'}^* \, \Pop.
  \end{equation}

We need to introduce the scale of oscillation of the phase function $\phi_o$, denoted $u_o$, to extract information from the equation above. We suppose then that $\phi_o(\mathbf{u}_o)$ is a function of $\vuo/u_o$. 
Since $ E_1^3$ is localized at the scale $\ell_1$, $K$ localizes $\vuo$ around $\vuo'$ at the scale $\ell_1^{-1}$. Hence, for a nearly widefield imaging scenario when $\ell_1$ is large compared to $u_o^{-1}$, $K(\vuo' - \vuo,0)$ is narrow, and we can replace $\CRs{\vuo'}^* \, \Pop$ by $\CRs{\vuo}^* \, P_o^*(\vuo)$, resulting in 
  \begin{equation}
    D D^*(\vuo,\vuo') = |\CRs{\vuo}|^2 \, K(\vuo' - \vuo,0),
  \end{equation}
  making $D D^*(\vuo,\vuo')$ nearly diagonal. In that case it is not possible to extract the aberrations from the distortion operator. We need then the illumination to be sufficiently focused, namely satisfying $u_o^{-1}  \geq \ell_1$ (but still satisfying $\ell_0 \ll \ell_1$), to extract the aberrations. In that case, $D D^*(\vuo,\vuo')$ is more extended around the diagonal and we have
  \begin{eqnarray*}
    \int D D^*(\vuo,\vuo') P_o(\vuo')d^2 \vuo' &=& \Po \,  \CRs{\vuo} \, \int K(\vuo' - \vuo,0) \,  \CRs{\vuo'}^* \,d^2\vuo'\\
    &\simeq & \Po \, |\CRs{\vuo}|^2  \int K(\vuo',0) \, d^2\vuo',
  \end{eqnarray*}
  where we used the string of inequalities $u_o^{-1}  \geq \ell_1 \gg \ell_0$ and the change of variables $\vuo' -\vuo \to \vuo'$. Since $\CRs{\vuo}$ varies at the scale $\ell_1^{-1}$, which is much greater than that of $\Po$, the function $\CRs{\vuo}$ is nearly constant compared to $\Po$ and as a consequence  $\Po$ can be separated from $|\CRs{\vuo}|^2$ and is approximately an eigenvector. Now, for $\Po$ to be estimated for all $\vuo$, the largest wavenumber that can be measured by the experimental system, denoted $u_m$, has to be much less than $\ell_0^{-1}$ since $|\CRs{\vuo}|^2$ becomes small when the length of $\vuo$ is greater than $\ell_0^{-1}$. In that scenario, $D D^*$ is approximately of rank one and the leading eigenvector of $D D^*$ gives an estimation of $\Po$. 

  In the opposite situation where $\ell_1 \ll \ell_0$, the analysis is similar as above with $\chi^{(3)}$ and $E_1^3$ swapped. Indeed, $\mathbf{x'}$ is now localized around $\mathbf{x}$ in $\CRc$, which becomes $\chi^{(3)*}(\mathbf{x})$, and changing variables as $\mathbf{x'}\to \mathbf{x}-\mathbf{x}_s$, we find
\begin{equation} \label{gammasup}
    \Gamma(\vuo,\vuo')  = \left(\int |\chi^{(3)}(\mathbf{x})|^2 e^{-i 2 \pi \mathbf{x} \cdot \Delta \mathbf{u}} \dx \right)\, \int K(\vuo' - \vuo, \v x'-\mathbf{x}_s)\,e^{i 2 \pi \mathbf{u}_0' \cdot (\mathbf{x}'-\mathbf{x}_s)} \dxp.
\end{equation}
The first term on the right above localizes $\Delta \mathbf{u}$ around 0 at the scale $\ell_0^{-1}$, while the integral of $K$ can be recast as
$$
 \hat{E}_1^3(\vuo)  (\hat{E}_1^3(\vuo'))^*
 $$
 where $\hat{E}_1^3$ is the Fourier transform of $E_1^3$. Based on the analysis for the case $\ell_1 \gg \ell_0$, the aberrations can then be reconstructed provided $u_o^{-1} \geq \ell_0$, namely the support of $\chi^{(3)}$ is sufficiently small compared to $u_o^{-1}$. The condition $u_m \ll \ell_1^{-1}$ is also necessary to estimate $\Po$ for all $\vuo$ since $\hat{E}_1^3(\vuo)$ becomes small when the length of $\vuo$ is greater than $\ell_1^{-1}$.

When $\ell_0$ corresponds to the scale of oscillation of $\chi^{(3)}$, we rewrite $\chi^{(3)}$ as
$\chi^{(3)}(\mathbf{x})=\varphi(\mathbf{x}/\ell_\varphi) e^{i 2 \pi \mathbf{u}_\chi \cdot \mathbf{x}}$ where $\varphi$ is an envelope function and the length of $\mathbf{u}_\chi$ is $\ell_0^{-1}$. Here $\ell_0 \ll \ell_\varphi$, and we assume $\ell_\varphi \gg \ell_1$. In that case, a short calculation shows that $\Gamma$ is as in (\ref{gammasup}) with $\chi^{(3)}$ replaced by $\varphi$ and the integral of $K$ becomes
$$
 \hat{E}_1^3(\vuo-\mathbf{u}_\chi)  (\hat{E}_1^3(\vuo'-\mathbf{u}_\chi))^*.
 $$
 The correlation matrix is not too narrow when $u_o^{-1} \geq \ell_\varphi$. The term $\Po$ has then to be separated from $\hat{E}_1^3(\vuo-\mathbf{u}_\chi)$, which is possible when $u_o^{-1} \gg \ell_1$, and  $\Po$ can be estimated for all $\vuo$ provided $\ell_1^{-1} \gg u_m$ and the length of $\mathbf{u}_\chi$, namely $\ell_0^{-1}$, is less than $u_m$.
 
 In the scenario where  $\ell_0$ and $\ell_1$ are comparable, the expression of $\Gamma$ does not simplify, but the kernel $K$ still does localize $\vuo$ around $\vuo'$ at the scale $\ell_1^{-1}$. Hence, as before, the aberrations can be reconstructed when the illumination is focused enough so that $u_o^{-1} \geq \ell_1$.



\begin{figure*}[hbt!]
    \centering
    \includegraphics[width=0.6\textwidth]{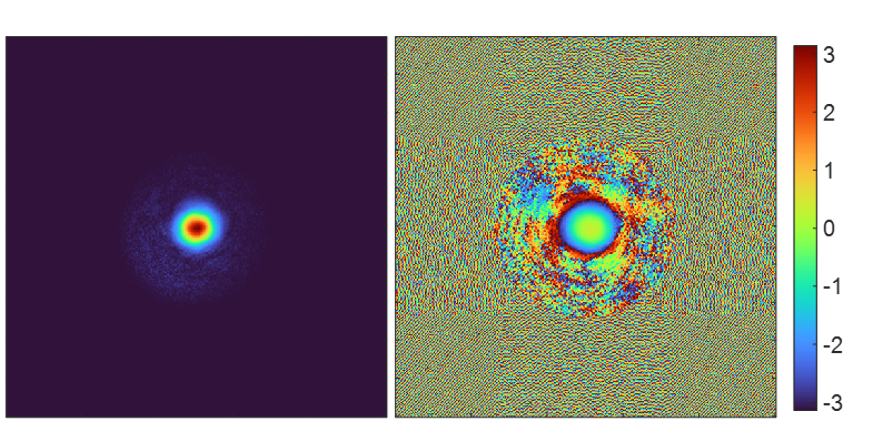}
    \caption{The amplitude (left) and phase (right) of the SVD first left singular value of the NDO. The images shown are in the spatial frequency domain with the circular diameter being approximately the spatial frequency support of the collection objective.  }
    \label{fig:first_singular_value}
\end{figure*}

The imaging system's pupil phase distortions are estimated from the first left singular value from the SVD of the mixed basis NDO, D($\vuo,\vxs$), which is the eigenvector of the correlation of the NDO, $D D^*(\vuo,\vuo')$. This is particularly the case in our setup when the illumination is stationary relative to the sample, making our the detection scheme isoplanatic. An example of the amplitude and phase of the SVD first left singular value of the NDO is shown in \rfig{first_singular_value}.

Our experimental setup allows us to directly measure the nonlinear distortion operator, perform a Fourier transform to get the mixed basis NDO, and estimate the pupil phase distortions and subsequently correct for those distortions. However, as pointed out in the main text, correcting for the dominant phase aberrations that occur from phase jumps in hologram-to-hologram measurements is necessary before using the NDO aberration correction.

\subsection{Optimization algorithm for spatial phase estimation and correction}

The phase of each hologram recorded from different spatial positions has a random phase offset, $\phir$, relative to each other due to random phase fluctuations between measurements. This random phase difference between the holograms of each $\vxs$ position in the transmission matrix prevents a good reconstruction of the THG field across the FOV. It is necessary to correct for these random phase fluctuations to coherently combine the THG fields. A brute force optimization approach can correct the phase by applying a piston phase shift to each reconstructed THG field in transmission matrix and maximizing the integrated intensity of the transmission matrix. Maximizing the intensity serves as a good parameter to optimize because when the phase offset of each field is corrected, adjacent fields will constructively combine and cause the intensity of the transmission matrix to increase. However, this brute force approach becomes computationally time consuming for large data sets. Therefore, we utilize a recently created adaptive phase algorithm for optimizing the phase matching of reconstructed THG fields in the transmission matrix. 

Dynamic adaptive scattering compensation holography (DASH) was originally designed for applying phase masks to a spatial light modulator (SLM) to optimize beam focusing through scattering media \cite{May2021}. The algorithm finds the optimal phase mask to project onto the SLM by continuously updating and improving the mask with every iteration. Here, we modify the DASH algorithm to apply a basis of digital phase masks to operate on the transmission matrix, rather than an SLM. The DASH phase optimized transmission matrix is defined as the original transmission matrix by the phase correction,
\begin{equation}
    T^c(\vxo,\vxs)  = \Txx \, \exp[- i \, \phi_r(\vxs)].
\end{equation}
The corrected image is then estimated by the computation
\begin{equation}
    I^c(\vxo) = \left| \int T^c(\vxo,\vxs) \, \dxs \right|^2.
\end{equation}

The phase correction pattern $\phir$ calculated with our modified DASH algorithm is defined by 
\begin{equation}
    \phir = \mathrm{angle}(C_{I,N}),
\end{equation}
where $C_{I,N}$ is the final correction mask after the iterative optimization and $N$ is the total number of recorded holograms, (i.e., the total number of $\vxs$ positions). A modulated wavefront mode, $M_n$, a phase step, $\phi_p$, and a reference field $C_{i,n}$ contribute to constructing the phase pattern 
\begin{equation}
    \Phi_{i,n,p} = \mathrm{angle} \left( \sqrt{1-f} \, \frac{C_{i,n}}{|C_{i,n}|}  + \sqrt{f} \, e^{i (M_n+\phi_p)} \right),
\end{equation}
where the subscript $i$ is the iteration number, $n$ is the mode number, $p$ is the phase step which is set to be $\phi_p = p(2\pi/P)$. We chose $P=5$, as recommended in the original DASH work, such that the $p=[0, 2\pi/5, 4\pi/5, 6\pi/5, 8\pi/5]$. The value of $f=0.3$ is chosen under recommendation of the authors of the original DASH work. The reference field for the first iteration is set to zero. The phase pattern is then applied the transmission matrix to maximize the cost function 
\begin{equation}
    J = \int I^c(\vxo) \, \dxo,
\end{equation}
where
\begin{equation}
    I^c(\vxo) = \left| \int T(\vxo,\vxs) \, \exp \left( i \, \Phi_{i,n,p} \right) \, \dxs \right|^2,
\end{equation}
The algorithm undergoes iterations to continually update the reference field 
\begin{equation}
    C_{i,n} +_1 = C_{i,n} + a_n\exp[i(M_n)]
\end{equation}
determined by an amplitude weighting function $a_n$. The weighting function is defined as

\begin{equation}
    a_n = \sum_p \frac{\sqrt{J}\exp(i\phi_p)}{P} 
\end{equation}

such that the phase correction mask is built from the basis set of modulated wavefront modes $M_n$ and the weighting function gives modes that increase the cost function amplitude, and thus, more contribution to the final phase correction mask $C_{I,N}$. After several iterations (typically 7 iterations for N=676 modes) the reference field converges to the final field $C_{I,N}$, which maximizes $J$.  

The result is a phase correction pattern that, when applied to the transmission matrix, maximizes the intensity of the transmission matrix, and thus corrects for the hologram-to-hologram phase distortions, $\phir$. The spatial overlap of each hologram provides the constructive/destructive interference necessary to optimize. Without spatial overlap, the phase of the holograms become independent of one another. The modified DASH algorithm provides a fast-converging optimization ($<$10 minutes) for moderately sized data ($\sim$ 4GB).  

Our modified DASH algorithm is used here for correcting phase differences between the signal and reference field measured for each $\vxs$ position. However, the pupil plane distortions still exist. We correct pupil plane phase aberrations through the distortion operator (DO) approach we have adapted for the THG transmission matrix.


\end{document}